\documentclass[final,authoryear,5p]{elsarticle}

 \usepackage{graphicx}

\usepackage{amssymb}

\usepackage{units}
\usepackage{upgreek}
\usepackage[latin1]{inputenc}

\usepackage[ps2pdf,%
a4paper=true,%
breaklinks=true,%
colorlinks=true,%
pdfauthor={Daniel Hampf et al.},%
pdftitle={Measurement of night sky brightness in southern Australia}%
]{hyperref}

\journal{Advances in Space Research}

\biboptions{authoryear}

\begin{document}

\begin{frontmatter}

\title{Measurement of night sky brightness in southern Australia}

\author[HH,Ad]{Daniel Hampf\corref{cor}}
\cortext[cor]{Corresponding author}
\ead{daniel.hampf@physik.uni-hamburg.de}
\author[Ad]{Gavin Rowell}
\author[Ad]{Neville Wild}
\author[Ad]{Tristan Sudholz}
\author[HH]{Dieter Horns}
\author[HH]{Martin Tluczykont}
\address[HH]{Department of Physics, University of Hamburg, Luruper Chaussee 149, 22761 Hamburg, Germany}
\address[Ad]{School of Chemistry \& Physics, University of Adelaide, Adelaide 5005, SA, Australia}

\begin{abstract}
Night sky brightness is a major source of noise both for Cherenkov telescopes as well as for wide-angle Cherenkov detectors. Therefore, it is important to know the level of night sky brightness at potential sites for future experiments.

The measurements of night sky brightness presented here were carried out at Fowler's Gap, a research station in New South Wales, Australia, which is a potential site for the proposed TenTen Cherenkov telescope system and the planned wide-angle Cherenkov detector system HiSCORE.

A portable instrument was developed and measurements of the night sky brightness were taken in February and August 2010. Brightness levels were measured for a range of different sky regions and in various spectral bands. 

The night sky brightness in the relevant wavelength regime for photomultipliers was found to be at the same level as measured in similar campaigns at the established Cherenkov telescope sites of Khomas, Namibia, and at La Palma. The brightness of dark regions in the sky is about $2 \times 10^{12}$ photons/(s~sr~m$^2$) between $\unit[300]{nm}$ and $\unit[650]{nm}$, and up to four times brighter in bright regions of the sky towards the galactic plane. The brightness in V band is 21.6 magnitudes per arcsec$^2$ in the dark regions. All brightness levels are averaged over the field of view of the instrument of about $\unit[1.3 \times 10^{-3}] {sr}$.

The spectrum of the night sky brightness was found to be dominated by longer wavelengths, which allows to apply filters to separate the night sky brightness from the blue Cherenkov light. The possible gain in the signal to noise ratio was found to be up to 1.2, assuming an ideal low-pass filter.
\end{abstract}

\begin{keyword}
Observatories and site testing; Airglow and aurorae; Photometric, polarimetric, and spectroscopic instrumentation
\PACS{95.45.+i, 92.60.hw, 95.55.Qf}
\end{keyword}

\end{frontmatter}

\parindent=0.5 cm


\section{Introduction}
The term ``night sky brightness'' (NSB) refers to the residual light that is present in the night sky during dark, moonless nights. It poses a major source of noise for ground-based astronomical observations, and good astronomical sites are therefore characterised by their low level of NSB (``darkness'') \citep{1989PASP..101..306G}. 

In the visible light regime, the main contributions of NSB are from air-glow, direct and scattered starlight and zodiacal light. 
Cities, but also facilities like ports or mining sites, produce a high level of light pollution, and the NSB near such locations is often dominated by artificial light. The pollution can extend many kilometres beyond the boundaries of the city, as can be seen in the world-wide atlas of night sky brightness \citep{2001MNRAS.328..689C}. At good observatory sites, however, the contribution of artificial sources to the total NSB should be less than 1\%, and the atlas provides a good first check for appropriate sites. 

A comprehensive overview about sources and measurements of NSB is given in \citet{1998A&AS..127....1L}. In \citet{1998NewAR..42..503B} a long term measurement of the NSB levels at the La Palma observatory is presented, including a detailed study of the spectrum of NSB using a spectrograph mounted at one the the observatory's telescopes. In the preparation of the H.E.S.S. telescope system, two candidate sites in La Palma and in Namibia have been examined in respect of their NSB levels, using a photon counting photomultiplier and a portable telescope mount \citep{2002NIMPA.481..229P}. A similar measurement has been conducted by \citet{NSB_Mirzoyan1994} using the first HEGRA telescope at La Palma.

NSB levels are a function of many variables: The amount of NSB varies between the different regions of the sky, especially between observations of the galactic plane and away from the plane. The altitude of the observation site has an effect on the absorption and the scattering of starlight and air-glow.  It is also a function of time on the time-scale of years (due to the solar activity cycle, which influences the amount of air-glow) as well as hours, as the air-glow decreases during the night  \citep{1998NewAR..42..503B}. These effects have to be taken into account when comparing measurements from different sites and different times.

In this paper, the night sky brightness is examined in the light of its effects on air Cherenkov light detectors for gamma-ray astronomy. Cherenkov telescopes like MAGIC \citep{2008ApJ...674.1037A}, VERITAS \citep{2008ApJ...679.1427A} and H.E.S.S. \citep{2006A&A...457..899A} observe cosmic gamma-rays and charged cosmic rays using the air Cherenkov technique: The primary particles interact with atoms in the upper atmosphere and produce a cascade of secondary, still highly energetic particles (air shower). These particles (mainly the electrons) emit Cherenkov light, and the air shower can be observed by Cherenkov telescopes with a very fast and sensitive camera system. The reconstruction of the primary particle properties (direction, energy, and others) uses the image properties of the air shower and, in some systems, its time evolution. 

NSB adds a constant level of background into each pixel of the camera, and its fluctuations limit the sensitivity to very weak signals. The spectral region from $\unit[280]{nm}$ to about $\unit[500]{nm}$ is the most relevant source of noise, as both the Cherenkov light spectrum and the spectral sensitivity of the photomultiplier tubes used in Cherenkov telescopes peak in the blue.

NSB is an even greater concern for wide-angle Cherenkov detectors like AIROBICC \citep{1995APh.....3..321K}, TUNKA \citep{ptuskin:tunka2008} and the proposed \mbox{HiSCORE} detector \citep{2009arXiv0909.0445T}. These instruments do not use an imaging system but integrate light from a large solid angle (up to 1~sr), which leads to a much higher level of NSB compared to Cherenkov telescopes, where the solid angle of a single pixel is of $\mathcal{O}(\unit[10^{-5}]{sr})$. In its simplest form, these detectors consist of an array of large photomultipliers aligned towards zenith, where the spacing between the detectors can be up to a few hundred meters. The reconstruction of the primary particle properties is done using the lateral photon distribution at detector level, the arrival times of photons and the duration of the Cherenkov light signals \citep{2009arXiv0909.0663H}. 

In these detectors the photon intensity due to NSB is high enough to not only disturb the air shower reconstruction but also to produce a non-negligible current in the photomultiplier anode, which requires the operation of these photomultipliers at low gain.

This work presents NSB measurements conducted at the Australian research facility Fowler's Gap\footnote{\url{http://www.fowlersgap.unsw.edu.au/}} (New South Wales, coordinates: 31$^\circ$ 05' S, 141$^\circ$ 43' E)
in the context of a campaign to find a suitable site for the proposed Australian Cherenkov telescope system TenTen \citep{2008NIMPA.588...48R} as well as the planned wide-angle Cherenkov detector HiSCORE. 
The site is about hundred kilometres away from the nearest city (Broken Hill), and from maps presented in  \citet{2001MNRAS.328..689C} the artificial light pollution is expected to be negligible. 

Fowler's Gap has been identified as a location with low average cloud coverage using publicly available weather data (G. Thornton et al 2011, in preparation). Also, the property of the research facility provides sufficient space for a large detector field like HiSCORE. 

Fowler's Gap is at an elevation of about $\unit[180]{m}$, which is far lower than the detector sites examined by the groups cited above. Simulations indicate that a low altitude site is very well suited for detectors that are designed for the ultra high energy regime  ($> \unit[30]{TeV}$) of gamma-rays, like HiSCORE and TenTen. It is however not a priori clear how NSB varies with altitude, and all studies so far have concentrated on high altitude sites. This study is therefore an important step towards the evaluation of a low altitude site for Cherenkov astronomy. 

The measurements presented here were carried out for the summer sky (14$^\mathrm{th}$ and 15$^\mathrm{th}$ February 2010) and the winter sky, including the Galactic Centre region (12$^\mathrm{th}$ and 13$^\mathrm{th}$ August 2010). 

The detector used follows the concept outlined in \citet{2002NIMPA.481..229P} with a few modifications to obtain a better calibration and a wider coverage of wavelengths. The detector and its calibration are described in detail in section \ref{detector}. Section \ref{evaluation} outlines the mathematical procedures used to obtain the NSB levels from the measurements. Section \ref{NSB_levels} summarises the NSB levels measured in Fowler's Gap for various regions of the sky. The measured spectrum of NSB is presented in section \ref{spectra} and the use of band-pass filters to suppress NSB is discussed in section \ref{filters}. Finally, the implications of the measured light levels for the TenTen telescope system as well as for the HiSCORE detector are discussed in section \ref{implications}.

\section{The instrument and its calibration}
\label{detector}
\subsection{Set-up}
To measure the night sky brightness levels, a photomultiplier module with photon counting capability (Hamamatsu HC124-3) is used. The module incorporates a R268P bi-alkali photomultiplier tube (PMT), a Cockroft-Walton high-voltage generator, an amplifier, a pulse shaper and a discriminator into a single, rugged metal casing. The threshold is internally set to discriminate noise from the single photon events, and signals are given according to the TTL standard. The PMT electronics can handle rates up to about $\unit[1]{MHz}$, and the dark count rate is in the order of $\unit[200]{Hz}$. 

\begin{figure}[tb]
  \begin{center}
    \includegraphics[width=0.4\textwidth]{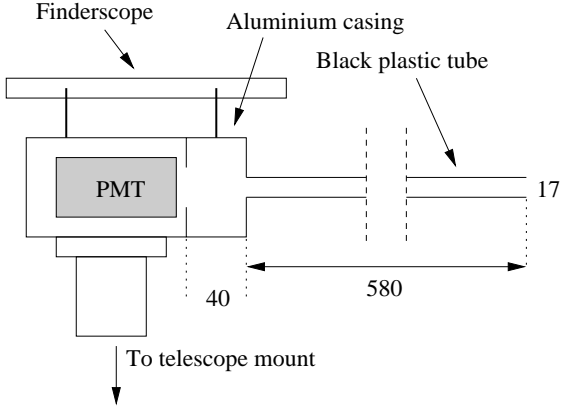} 
    \caption{Sketch of the set-up used}
    \label{set-up_sketch}
    \end{center}
\end{figure}

\begin{figure}[tb]
  \begin{center}
    \includegraphics[width=0.4\textwidth]{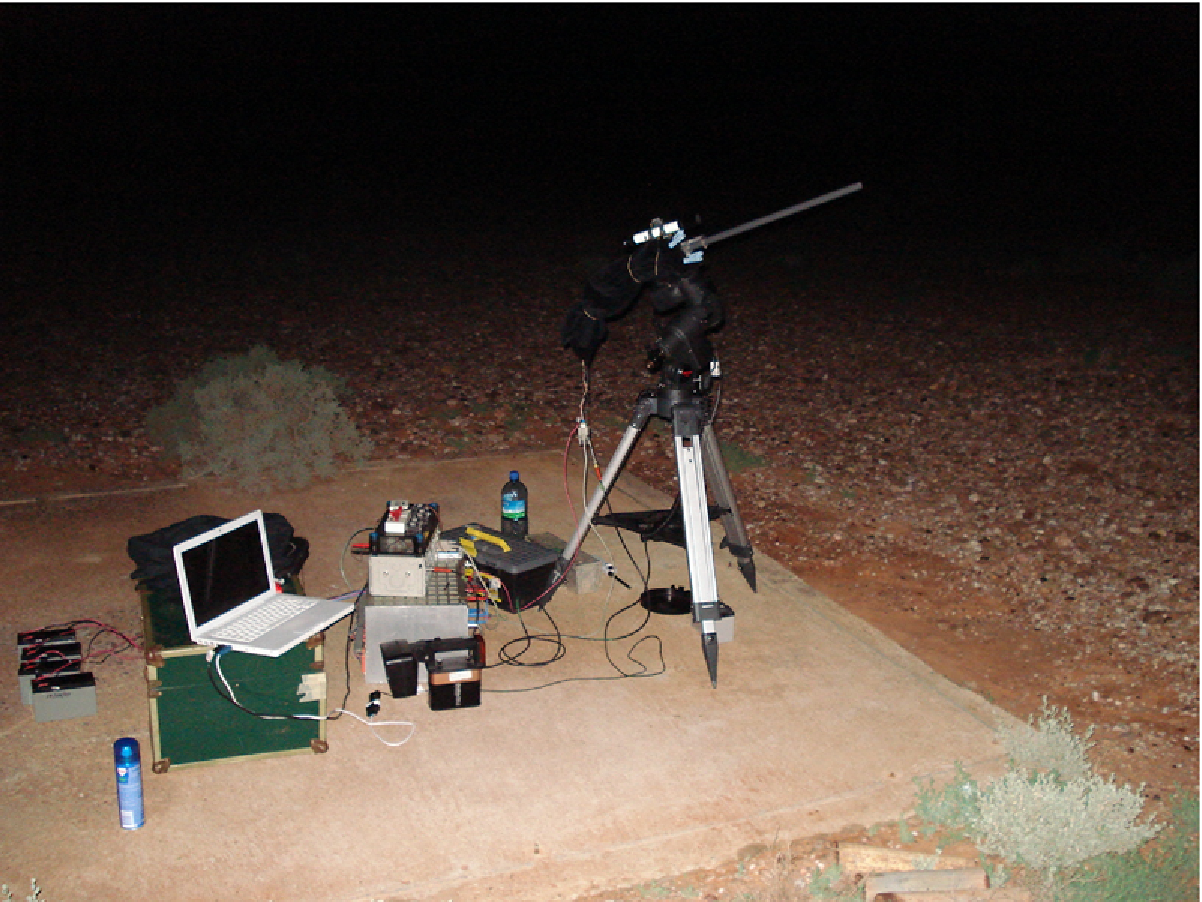} 
    \caption{The set-up during the measurements at Fowler's Gap}
    \label{set-up_pic}
    \end{center}
\end{figure} 

The PMT is fixed in a metal tube, which in turn is installed on a standard telescope mount (Meade LX D55). The telescope mount can be controlled by a hand-held controller or by a computer using a serial connection. A long plastic tube (inner diameter 17 mm, length 580 mm) is positioned in front of the PMT in order to reduce the rate of the NSB photons to rates between 50 kHz and several hundred kHz. The set-up is shown in figures \ref{set-up_sketch} and \ref{set-up_pic}.

The PMT signals are read by an analogue-digital converter (National Instruments USB-6008) which is connected via USB to a laptop and is being controlled by a LabView programme. Additionally, this programme controls the high voltage of the PMT and records the ambient temperature as well as the position of the telescope mount in equatorial coordinates. All data are logged into an ASCII file for later analysis. 

The supply voltages for the PMT module and the temperature readout electronics are generated by four 12 V rechargeable lead batteries.

\subsection{Effective solid angle}
\label{solid_angle}
In order to convert recorded photomultiplier count rates into a photon intensity in SI units, it is necessary to know the effective area and the effective solid angle of the device. 

As effective area, the area of the upper tube opening is used:
\begin{eqnarray}
   A  &=& (\unit[(8.38 \pm 0.02)] {mm}) ^2 \cdot \pi  \nonumber \\ 
       &=& \unit[(2.21 \pm 0.01) \times 10^{-4}] {m^2}  
  \label{area}
\end{eqnarray}

The effective solid angle can be determined (if the acceptance of the PMT is assumed to have no azimuthal dependence) by:
\begin{equation}
  \Omega = 2 \pi \int \! \epsilon(\theta) \sin(\theta) \, d\theta
  \label{solid_angle_eq}
\end{equation}
where $\epsilon(\theta)$ is the angle-dependent transmission through the plastic tube \citep{2002NIMPA.481..229P}. 
The transmission $\epsilon(\theta)$ is calculated by a ray-tracing algorithm, which contains the (a priori unknown) reflectivity of the black plastic tube as a free parameter. An analytical calculation has been performed for the case of no reflection in order to cross-check the ray-tracing simulation. In both cases the reflectivity has been assumed to be independent of angle and wavelength.

\begin{figure}[tb]
  \begin{center}
    \includegraphics[angle=270, width=0.4\textwidth]{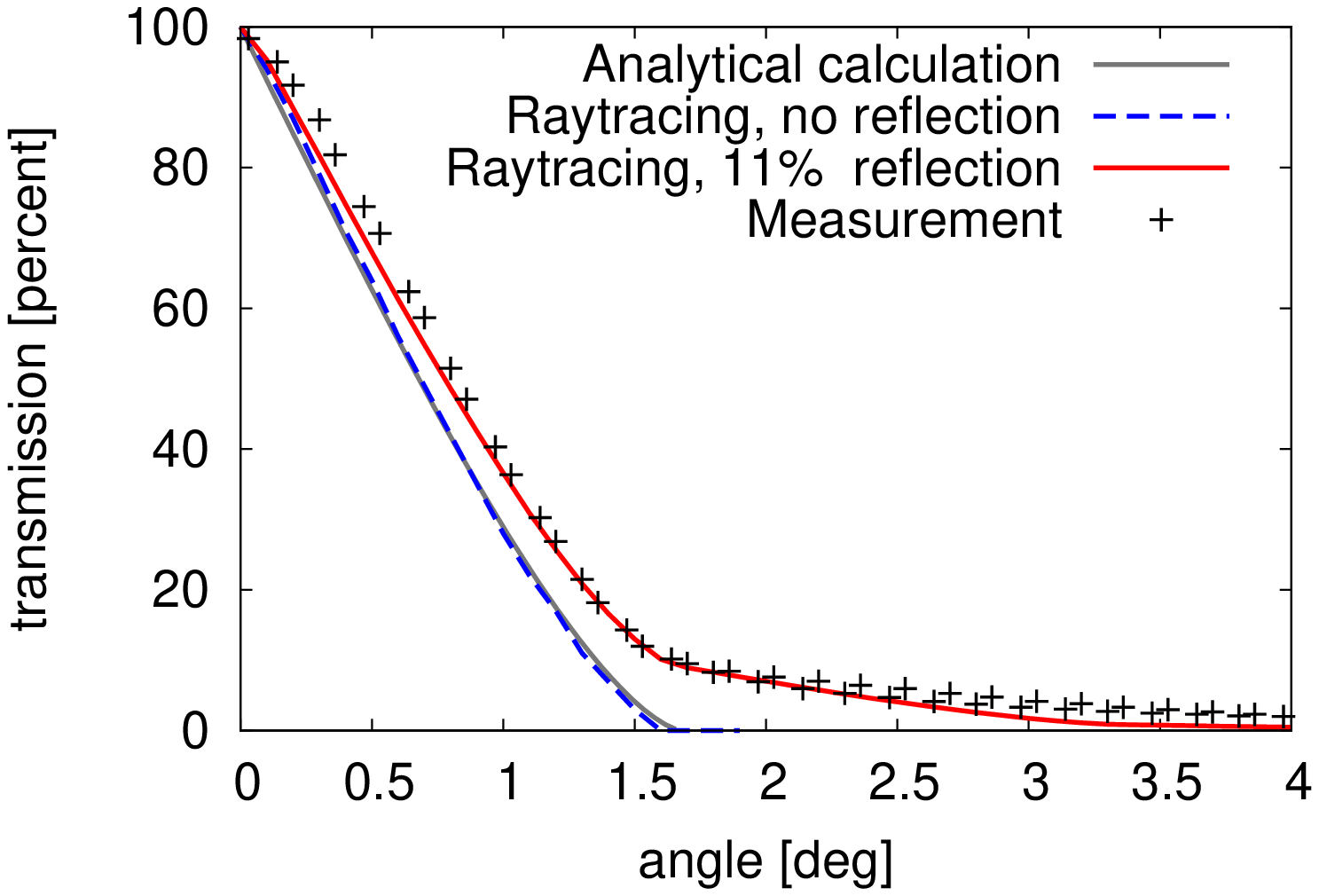} 
    \caption{The transmission of the plastic tube as determined by analytical calculation, ray-tracing simulation and measurement}
    \label{transmission_vs_angle}
    \end{center}
\end{figure}

In order to further cross-check the simulation and to obtain the reflectivity, $\epsilon(\theta)$ has also been determined experimentally. For this, the set-up was pointed towards a point-like light source in about twenty meters distance in a dark indoor corridor. Rays reaching the front opening of the plastic tube from that distance are nearly parallel, the maximum difference of angles is less than 3 arc-minutes. Using the telescope mount controls the tube can be pointed to various angles away from the light source, and for each angle the intensity of the transmitted light is recorded by the PMT module. The results of the calculation, the ray-tracing simulation and the measurements, all normalised to the intensity at parallel incident light, are shown in figure \ref{transmission_vs_angle}.

The analytical calculation and the simulation result for the case of no reflection agree very well. Simulations were carried out for different reflectivities in one percent steps, and the result of the simulation with 11\% reflectivity describes the experimental values best. Towards higher angles the measured transmissions are slightly above the ray-tracing values; this is mainly due to stray light being reflected into the device from the corridor walls and residual light in the corridor. At small angles the measured values are higher than expected as well. The reason for this is unknown; however, the contribution of these values to the effective solid angle is small.

A numerical integration of Eq.\ \ref{solid_angle_eq} for the simulation with 11\% reflectivity gives an effective solid angle of 
$\Omega = \unit[1.24 \times 10^{-3}] {sr}  $.
If the integration is performed using the measured values for the inner 1.5$^\circ$ and the ray-tracing results for the larger angles, the effective solid angle becomes
\mbox{$  \Omega =  \unit[1.29 \times 10^{-3}] {sr} $}.
For the following calculations a value of 
\begin{equation}
  \Omega  =  \unit[(1.29 \pm 0.05) \times 10^{-3}] {sr}  
  \label{solidangle}
\end{equation}
is used.

\subsection{Linearity and long-term stability}
In order to check the accuracy of the frequency measurement with LabView, several cross-checks with a stand-alone rate meter were performed, and no discrepancy was found up to $\unit[1]{MHz}$.

To check the linearity of the count rate to the incident light intensity, the device has been pointed at a bright region of the sky and several combinations of neutral density filters have been used to reduce the rate by known ratios. The deviations from linearity were within 4\%.

The stability has been tested in various long runs (12 to 24 hours) of the device in a dark room. A LED was switched automatically to several defined intensities in regular intervals and the count rate of the PMT module was recorded. Deviations throughout the night were less than 1\%, and deviations between different runs were smaller than 2\%.

\section{Evaluation of data}
\label{evaluation}

The night sky brightness values measured during this project will be given in the following ways:
\begin{itemize}
\item 
\textbf{Count rate} at PMT: This is the count rate that was recorded by the device during the measurements. It is given in order to allow direct comparison of results of future campaigns using the same device.

\item 
\textbf{Photoelectron intensity} at PMT: This value gives the above count rate in units of (s~sr~m$^2$)$^{-1}$, using the area and the solid angle given in Eq.\ \ref{area} and \ref{solidangle}. This can be used directly as input for detector simulations, assuming that the PMT used here and the ones used in the experiments have sufficiently similar spectral response and peak quantum efficiency. The systematic uncertainty of this value is caused by uncertainties in the solid angle, the effective area of the tube and slight changes in the dark count rate, and is estimated to be about 5\%.

\item
\textbf{Integral photon intensity} between $\unit[300]{nm}$ and $\unit[650]{nm}$: This value gives the intensity of photons incident at the PMT cathode in units of photons/(s~sr~m$^2$). For this, an effective quantum efficiency of the PMT has to be computed by convolving its wavelength dependent quantum efficiency $\epsilon_{\mathrm{PMT}}(\lambda)$ and the normalised spectrum of the night sky brightness $S(\lambda)$ \citep{2002NIMPA.481..229P}:
\begin{equation}
	\left< \epsilon_{\mathrm{PMT}} \right> = \int_{\lambda 1}^{\lambda 2}  \! S(\lambda) \epsilon_{\mathrm{PMT}}(\lambda) \, d\lambda
\end{equation}
Using the wavelength dependent quantum efficiency given by the manufacturer of the PMT and the night sky spectrum measured by \citet{1998NewAR..42..503B}, an effective quantum efficiency of 
\begin{equation}
	\left< \epsilon_{\mathrm{PMT}} \right> = 0.0935
	\label{QE}
\end{equation}
is found. It is estimated that this value has an uncertainty of about 10\%, resulting in an uncertainty of the integral photon intensity of 15\%.

While the integral photon intensity allows the easiest comparison to other studies taken with different instruments (e.g., CCD cameras at telescopes), it is subject to larger systematic uncertainties introduced by the uncertainties in the quantum efficiency and the assumed night sky spectrum.

\end{itemize}

The spectral measurements presented in section \ref{spectra} were done using a set of astronomical Johnson filters (U, B, V, and R). To get meaningful spectra, the recorded count rates $R$ have to be corrected for the different total transmission of the filters and the wavelength dependency of the PMT quantum efficiency, resulting in a differential photon intensity (unit: photons/(s~sr~m$^2$~nm)):
\begin{equation}
	\Phi = \frac{R}{A \, \Omega \, w_x}
	\label{spectral_rates}
\end{equation}
In this, $A$ und $\Omega$ are the area and the effective solid angle of the tube as defined in Eq.\ \ref{area}  and \ref{solidangle}, while $w_x$ is the effective wavelength window size of the filter x, defined as
\begin{equation}
	w_x = \int \! \epsilon_{\mathrm{PMT}}(\lambda) T_x(\lambda) \, d\lambda
	\label{sx}
\end{equation}
with the PMT quantum efficiency $\epsilon_{\mathrm{PMT}}(\lambda)$ and the transmission function of the filter $T_x(\lambda)$. The transmission functions were taken from the manufacturer's data sheets; a cross-check with a spectrograph gave consistent results.  
Table \ref{filter} shows the calculated values of $w_x$ for the four Johnson filters.

\begin{table}[tb]
\begin{center}
\begin{tabular}{ccc}
\hline
Filter		& Eff. wavelength midpoint &	$w_x$ 	\\
\hline
U 		& $\unit[365]{nm}$	& 	$\unit[9.00]{nm}$	\\
B 		& $\unit[445]{nm}$	&	$\unit[15.4]{nm}$	\\
V 		& $\unit[551]{nm}$	& 	$\unit[5.44]{nm}$	\\
R 		& $\unit[658]{nm}$	& 	$\unit[1.40]{nm}$	\\
\hline
\end{tabular}
\end{center}
\caption{The effective wavelength window size $w_x$ for the four Johnson filters that were used during the spectral measurements, calculated by Eq.\ \ref{sx}. The effective wavelength midpoints are taken from \citet{1998gaas.book.....B}.}
\label{filter}
\end{table}

A commonly used unit for the brightness of the night sky is \emph{magnitudes per arcsec$^2$} in the B and V band. 
The differential photon intensity can be converted to this unit using equations from \citet{1998A&AS..127....1L}:
\begin{eqnarray}
  \textrm{B band:} \hspace{3mm}\unit[4.81 \times 10^7]{\frac{\mathrm{photons}}{s\ sr\ m^2\ nm}} = \textrm{1 S$_{10}$ unit} \\ 
  \textrm{V band:} \hspace{3mm}\unit[3.27 \times 10^7]{\frac{\mathrm{photons}}{s\ sr\ m^2\ nm}} = \textrm{1 S$_{10}$ unit} \\
  \textrm{1 S$_{10}$ unit} 		= 27.28 \frac{\mathrm{mag}}{\mathrm{arcsec^2}}
\end{eqnarray}

\section{The brightness of the sky}

\begin{table*}[tb]
\begin{center}
\begin{tabular}{llcccccc}
\hline
Name		& Constellation	& RA		& Dec 		& Count Rate	& p.e.\ intensity  &  photon intensity &  \\
		&		& [hh:mm]	& [deg:mm]	&  [kHz]	&  [10$^{11}$ Hz] & [10$^{12}$ Hz]  \\	
			&			&		&		&	& $\pm 5\%$	& $ \pm 15\%$ 	\\
\hline
Dark region 1 (F)	& Hydra 	& 8:25 	& 0:00  	 & 55	& 1.9 	& 2.1  \\
Dark region 2 (F) 	& Antlia 	& 10:46 & -25:37 	& 48	& 1.7	& 1.8  \\
Dark region 3 (F)	& Lepus 	&  5:25	& -26:16 	& 47	& 1.7	& 1.8   \\
Dark region 4 (F)	& Corvus 	&  12:42& -16:50 	& 56	& 2.0 	& 2.1   \\

Dark region 5 (A)	& Sculptor	& 23:30	& -32:29  	& 57	& 2.0 	& 2.1   \\
Dark region 6 (A)	& Piscis Australis& 22:20& -24:00  	& 54	& 1.9 	& 2.0   \\
Dark region 7 (A)	& Virgo 	& 13:40	& -8:50  	& 76	& 2.7 	& 2.9   \\
Dark region 8 (A)	& Libra 	& 14:30	& -17:50  	& 71	& 2.5 	& 2.7   \\

From [1] & \multicolumn{3}{c}{\emph{- - - various - - -}} & n / a	&  2.3	& 2.4     \\
From [2] & \multicolumn{3}{c}{\emph{- - - various - - -}} & n / a	&	& 1.8     \\

\hline
Bright region 1	(F)	& Monoceros	& 7:25	& -10:36 	& 83	& 2.9 & 3.1    \\
$\upeta$-Carinae area (F) & Carina	& 10:44& -59:52	 	& 97	& 3.4	& 3.7    \\
LMC	(F)		& Dorado	& 5:23	& -64:44	& 100	& 3.5	& 3.8  \\
Crux region (F)		& Crux		& 11:22	& -64:34 	& 138	& 4.9	& 5.2  \\

GC region (A) & Sagittarius 		&  17:20  & -30:00 	& 189	& 6.6 & 7.1  \\

\hline

\end{tabular}
\end{center}
\caption{Night sky brightness levels observed in February (F) and August (A) 2010 from Fowler's Gap, and references from \citet{2002NIMPA.481..229P} ([1], Namibia and La Palma) and \citet{NSB_Mirzoyan1994} ([2], La Palma). The upper part of the table contains selected dark regions, the lower part brighter regions of the sky. The photoelectron and photon intensities are given per $\unit{sr}$ and $\unit{m^2}$ (see section \ref{evaluation} for definition of these numbers). LMC denotes the Large Magellanic Cloud and GC the Galactic Centre.}
\label{observations}
\end{table*}

\label{NSB_levels}
\subsection{Dark regions}

During the measurements in Fowler's Gap, the device was pointed at several selected dark regions in the sky. 
Each region was measured several times each night, and the results were averaged. The differences between the measurements were less than 5\%, and probably mostly due to the different zenith angles of the individual measurements (unfortunately, not enough data could be collected during this campaign to evaluate the zenith angle dependence systematically).

The upper part of table \ref{observations} gives the coordinates and the measured brightness levels at several dark regions that were observed during the two campaigns in February and August 2010. Most values found here are slightly lower than the one given in \citet{2002NIMPA.481..229P} and higher than in \citet{NSB_Mirzoyan1994}, but consistent with both within the systematic uncertainties. 
Further comparisons with other measurements, which were taken in certain photometric bands, are given in section \ref{spectra}.

The dark regions 7 and 8 (Virgo and Libra constellations) are significantly brighter than the rest of the dark regions, even though they are well away from the galactic plane and other bright regions. However, since they were close to the horizon during the time of observation and their position on the sky is coincident with the ecliptic, it is possible that the higher photon intensity is at least partly due to zodiacal light. Comparing with measurements of the brightness of zodiacal light, e.g.\ by \citet{2002ApJ...571...85B}, one finds that the observed increase of brightness is in the expected order of magnitude.


\subsection{Bright regions}

\begin{figure}[tb]
  \begin{center}
    \includegraphics[angle=270, width=0.5\textwidth]{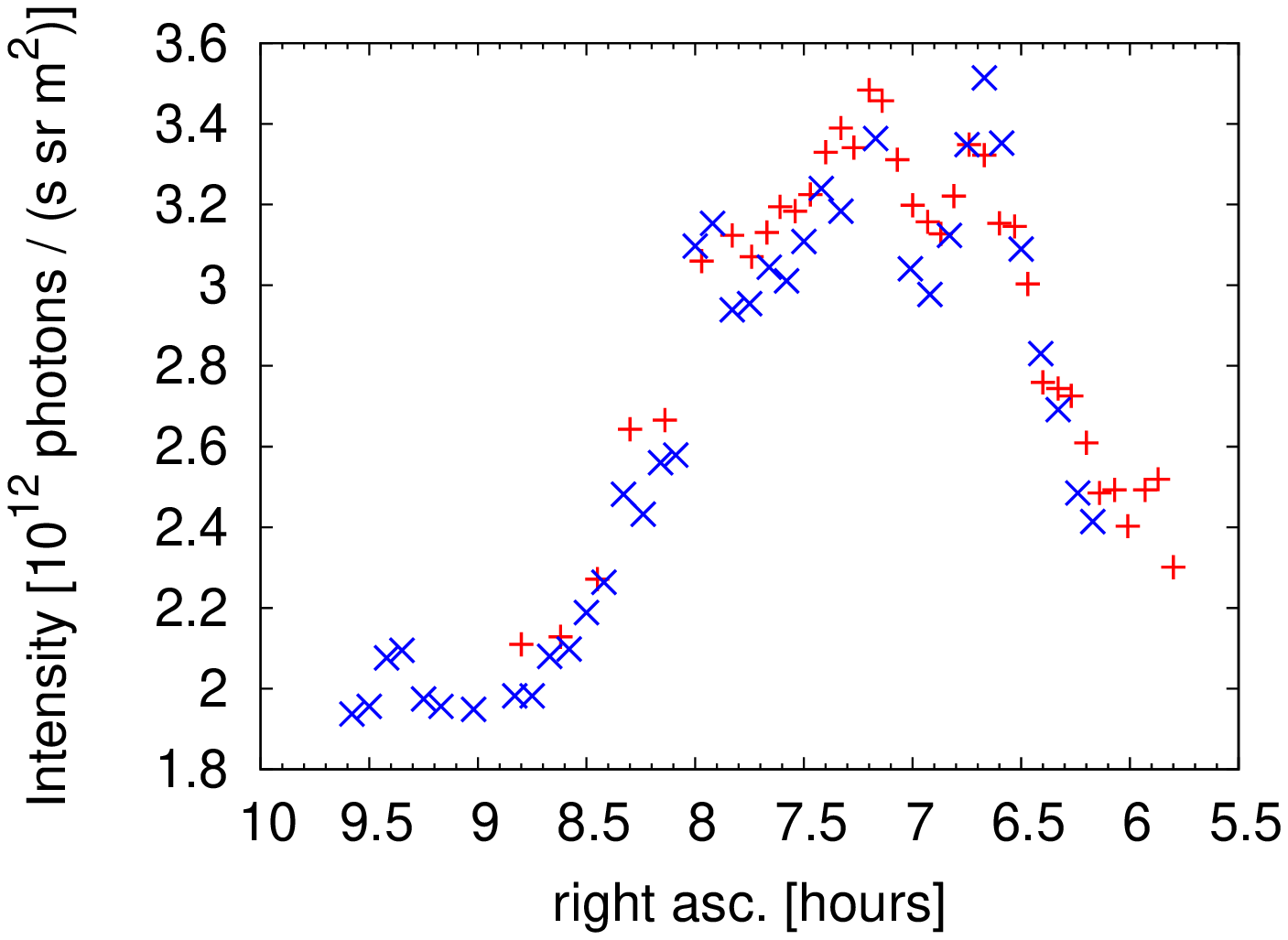} 
    \caption{Scan along a declination of -22$^\circ$ through the Canis Major region. The different markers present measurements from two different nights during the February campaign.}
    \label{canis_scan}
    \end{center}
\end{figure}

\begin{figure}[tb]
  \begin{center}
    \includegraphics[angle=270, width=0.5\textwidth]{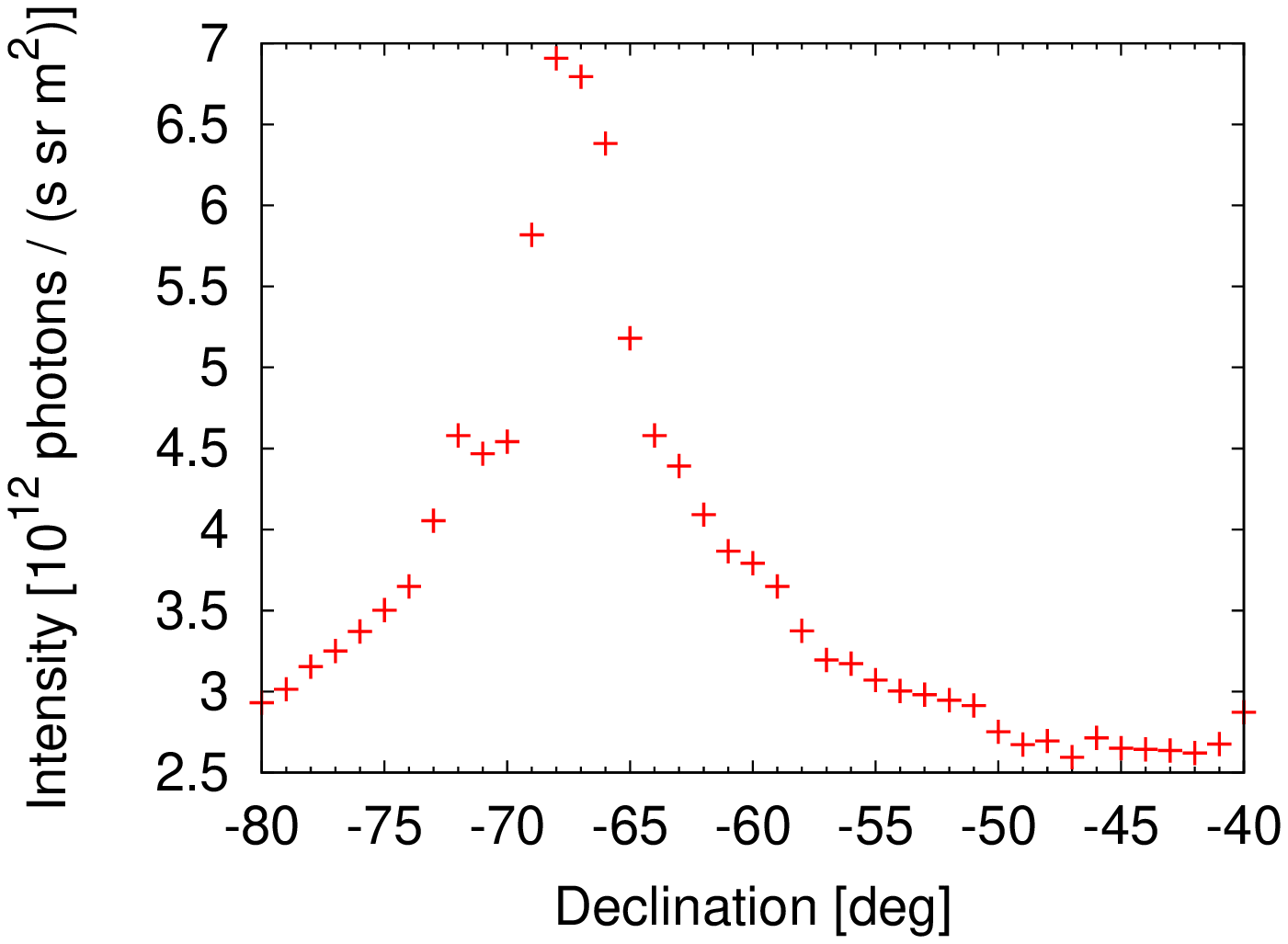} 
    \caption{Scan along the right ascension of 11:30 hours through the Southern Cross region, taken in February 2010.}
    \label{crux_scan}
    \end{center}
\end{figure}

\begin{figure}[tb]
  \begin{center}
    \includegraphics[angle=270, width=0.5\textwidth]{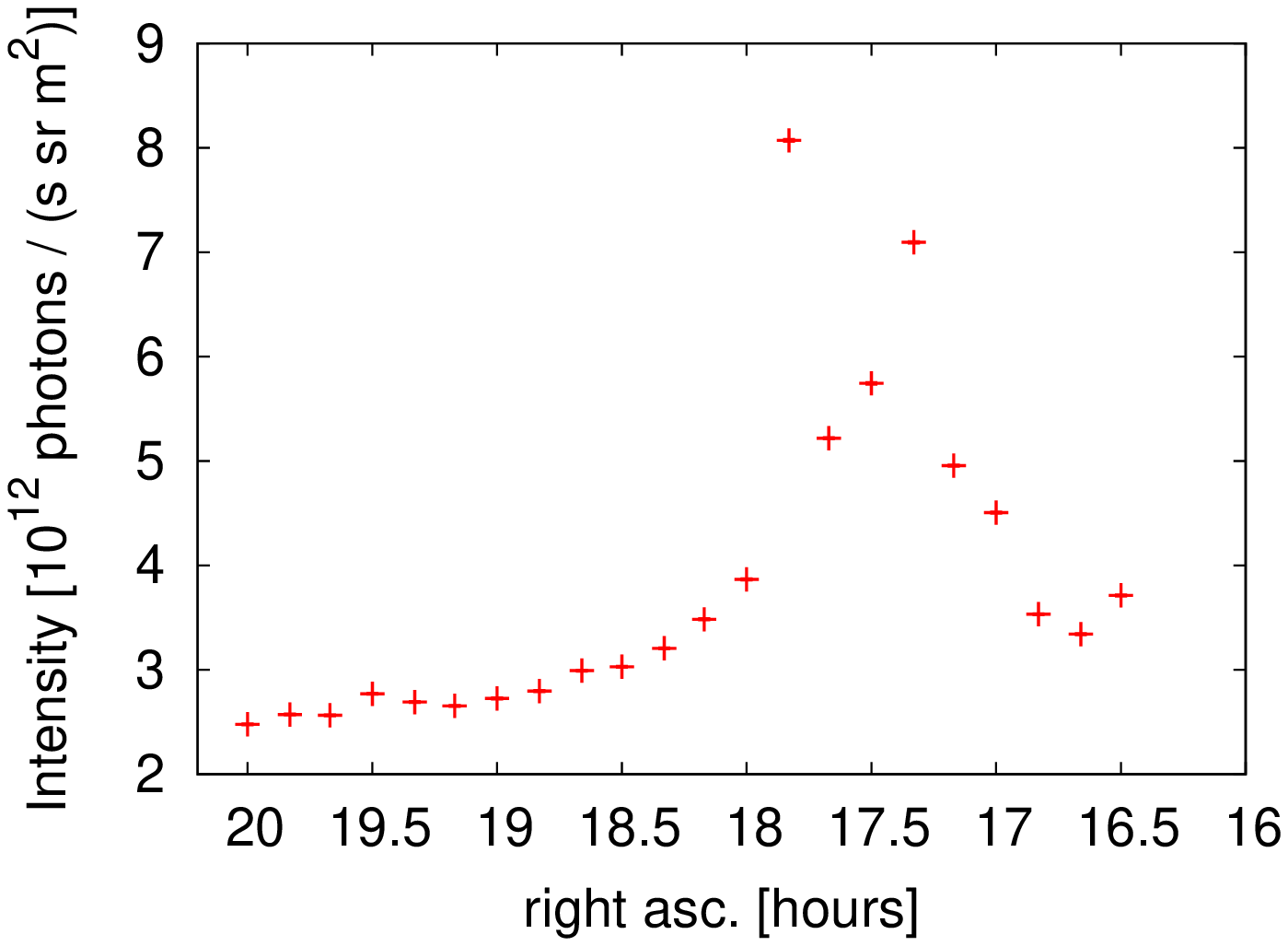} 
    \caption{Scan along the declination of -30$^\circ$ through the Galactic Centre region, taken in August 2010.}
    \label{GC_scan}
    \end{center}
\end{figure}

As many known and potential high energy gamma-ray sources are located in brighter regions of the sky like the galactic plane, the night sky brightness levels in these regions are of great interest for this study. The lower part of table \ref{observations} shows some measurements from brighter regions of the sky.

Figure \ref{canis_scan} shows two scans across the plane in the Canis Major region, one of the darker regions of the galactic plane. The two scans, which were taken along a declination of -22$^\circ$ in two successive nights, are in reasonable agreement. The maximum photon intensity in this scan is about $\unit[3.5 \times 10^{12}]{photons/(s~sr~m^2)}$, almost twice as bright as in the darkest regions of the sky.

Figure \ref{crux_scan} shows a scan across the plane in the region of the Crux, one of the brighter regions in the galactic plane. The scan was taken along a right ascension of 11:30 hours, and the maximum photon intensity is about $\unit[7 \times 10^{12}]{photons/(s~sr~m^2)}$, twice as high as in the Canis Major area and almost four times as high as in the darkest regions of the sky.

The situation is similar in the Galactic Centre / Sagittarius region, one of the brightest regions in the sky, a scan of which is shown in figure \ref{GC_scan}. The very bright outlier at 17.8~hours is probably due to a single bright star, while the actual peak of the scan is found at 17.3~hours with a brightness of $\unit[7 \times 10^{12}]{photons/(s~sr~m^2)}$. This confirms the measurement of \citet{2002NIMPA.481..229P} who found that the Galactic Centre region is up to four times brighter than the darkest regions of the sky.

The statistical errors in the data points of figures \ref{canis_scan} to \ref{GC_scan} are negligible. The systematic error is assumed to be 15\%, mainly due to uncertainties in the instrument calibration (see section \ref{evaluation}).


\section{Spectral composition of the night sky brightness}
\label{spectra}

\begin{figure}[tb]
  \begin{center}
    \includegraphics[angle=270, width=0.5\textwidth]{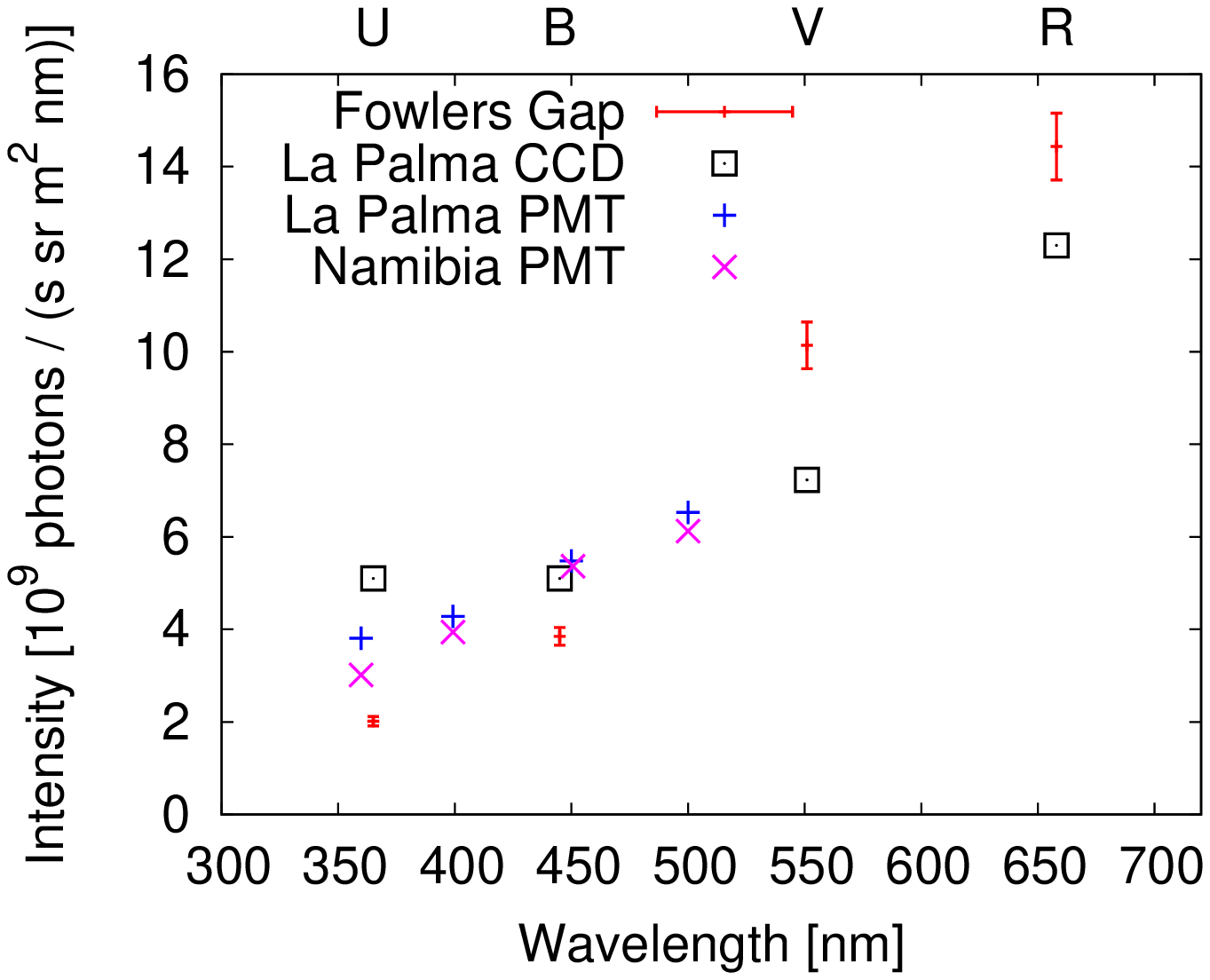} 
    \caption{Averaged spectra of dark regions 1, 2, 3, 6, and 7 as measured from Fowler's Gap with astronomical Johnson filters (U, B, V, and R bands) and measurements from \citet{2002NIMPA.481..229P} ("PMT") and \citet{1998NewAR..42..503B} (''La Palma CCD``). Statistical uncertainties are indicated by errorbars, systematic uncertainty is 15\% (see section \ref{evaluation}).}
    \label{spectra_overview}
    \end{center}
\end{figure} 

\begin{figure}[tb]
  \begin{center}
    \includegraphics[angle=270, width=0.5\textwidth]{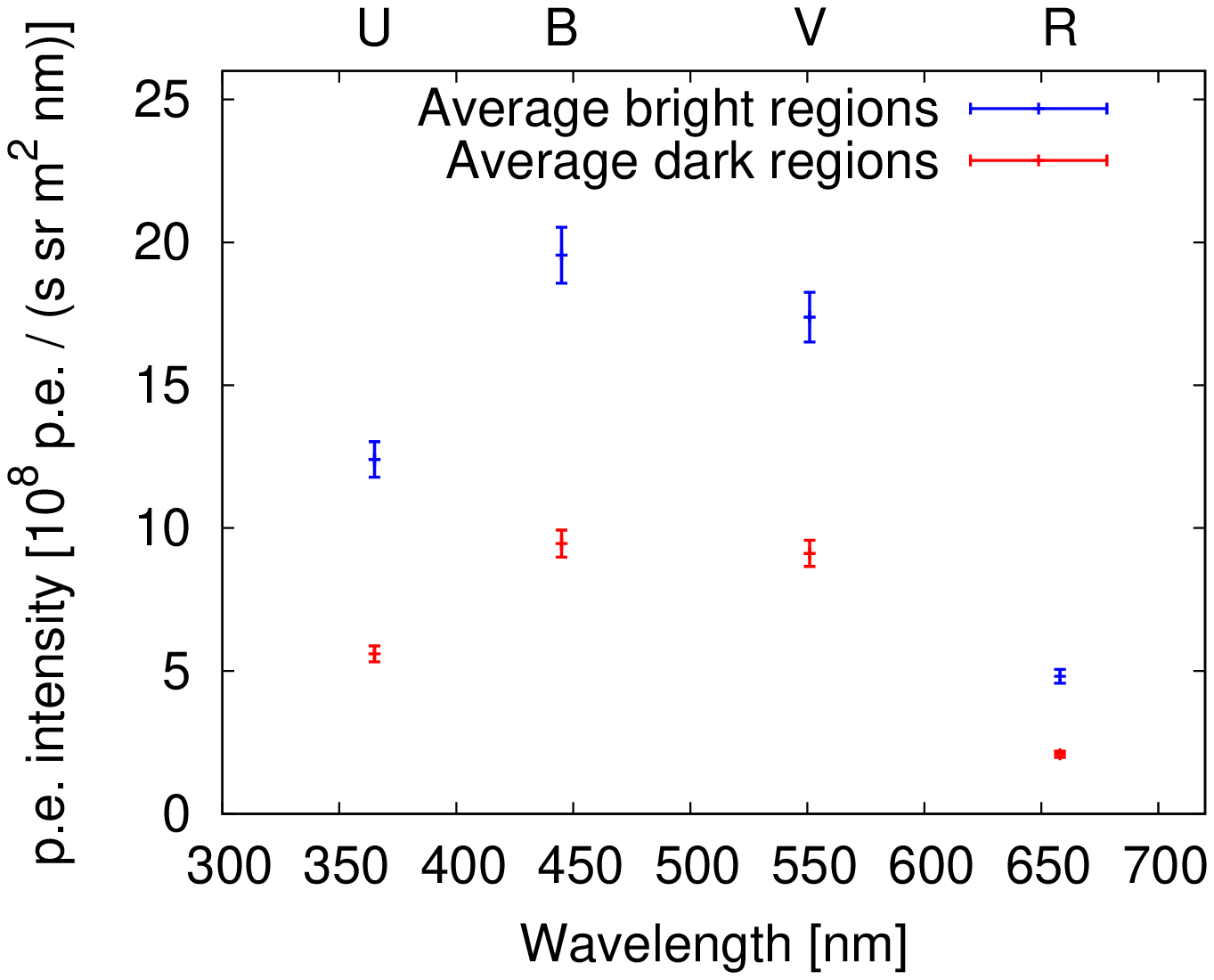} 
    \caption{Spectra of dark and bright regions as measured in Fowler's Gap, without unfolding the PMT response (i.e., the light levels actually seen by the PMT). Statistical uncertainties are indicated by errorbars, systematic uncertainty is 15\% (see section \ref{evaluation}).}
    \label{raw_spectra}
    \end{center}
\end{figure}

The spectral composition of the NSB has been examined by using standard astronomical Johnson filters in front of the entry window of the set-up. The measured rates in U, B, V, and R band are converted to a differential photon intensity at the effective mid-wavelength of the filters using Eq.\ \ref{spectral_rates}.

The averaged spectrum of the dark regions 1, 2, 3, 6, and 7 (as defined in table \ref{observations}) is shown in figure \ref{spectra_overview} along with the results of other measurements.
In \citet{2002NIMPA.481..229P} the spectra were measured with a PMT and narrow bandpass filters (FWHM between $\unit[3.5]{nm}$ and $\unit[4.3]{nm}$) and averaged over several regions near zenith. The campaign took place in June and July 2000, in a phase of high solar activity.
The spectrum cited from \citet{1998NewAR..42..503B} was measured with a CCD camera at the William Herschel Telescope at La Palma near a minimum of solar activity. Both were taken at high altitudes, $\unit[1800]{m}$ above sea level in Namibia and $\unit[2200]{m}$ a.s.l.\ at La Palma, while the measurements from Fowler's Gap were taken at $\unit[180]{m}$ a.s.l., and during a medium solar activity.

While the general shape of all spectra in figure \ref{spectra_overview} is similar, the night sky at Fowler's Gap is brighter at long wavelengths and is slightly darker at the blue end of the spectrum compared to the La Palma data. Since the solar activity was higher during the measurements at Fowler's Gap than during the measurements at La Palma, a higher level of NSB is expected due to increased air-glow. This seems to be the most natural explanation for the higher photon intensities observed in V and R band. On the other hand, the lower altitude of Fowler's Gap as compared to all other shown measurements should lead to stronger absorption of night sky brightness, especially at shorter wavelengths, which can explain the lower intensities seen in U and B band.

PMTs are much less sensitive at longer wavelengths than in the blue, so that the light in V and especially in R band does not contribute as much to the overall brightness seen by the PMT as the U and B bands. Figure \ref{raw_spectra} shows the spectra before the unfolding of the PMT response, i.e.\ the light levels actually seen by the PMT, for an average of dark regions (the same as in the plot of figure \ref{spectra_overview}) and bright regions (the Crux region, the LMC and the Monoceros region as defined in table \ref{observations}). The highest contributions come from the B and V band, which are of roughly the same strength even though the photon intensity in V band is almost three times higher than in B band (see figure \ref{spectra_overview}). There is less contribution from the U band and even less from the R band, even though the NSB is dominated by long wavelengths. There is no significant difference in the shape of the spectrum between dark and bright regions of the sky.

Table \ref{band_rates} shows the measured brightness in B and V band in different units. The magnitude in V band of a dark region of the sky is often used as a measure for the darkness of an astronomical site. A value of $\unit[21.6]{mag} / \unit{arcsec^2}$ can be considered to indicate a good dark site \citep{1989PASP..101..306G} and compares to the most renowned observatory sites of the world: Measured values for the European Southern Observatory in Chile are between 21.69 and 21.91, and for the McDonald Observatory in the United States between 21.54 and 21.92 \citep{1998A&AS..127....1L}. In those measurements all stars brighter than magnitude 13 were excluded, which was not possible in the measurements at Fowler's Gap, so that a comparable measurement would give an even (slightly) better value at Fowler's Gap.

\begin{table*}[t]
\begin{center}
\begin{tabular}{lccc}
\hline			
			& $10^9$ photons / (s sr m$^2$ nm) & S$_{10}$ units & $\unit{mag} / \unit{arcsec^2}$ \\
\hline
Dark regions, B	band     & 3.85		& 80	& 23.0  \\
Dark regions, V band 	& 10.1		& 310	& 21.6  \\
Bright regions, B band	& 7.97		& 166	& 22.2 \\
Bright regions, V band	& 19.4		& 592 	& 20.9 \\
\hline
\end{tabular}
\end{center}
\caption{Measured NSB levels at Fowler's Gap in B and V band in different units.}
\label{band_rates}
\end{table*}

\section{Using filters to improve signal to noise ratio}
\label{filters}
\begin{figure}[t]
  \begin{center}
    \includegraphics[angle=270, width=0.5\textwidth]{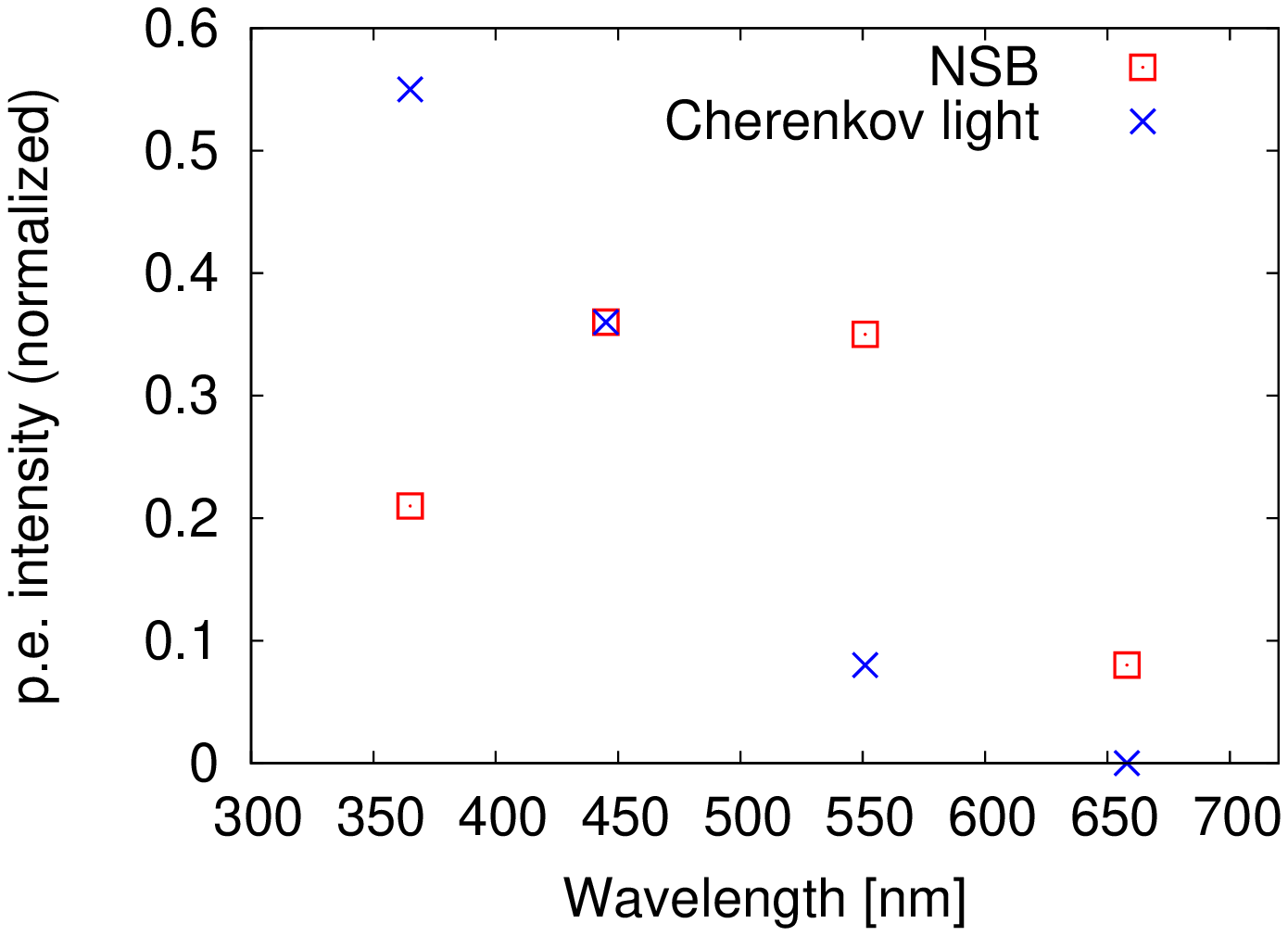} 
    \caption{Normalised spectrum of the dark regions (the same as in figure \ref{spectra_overview}), and simulated Cherenkov light intensities in the four used bands (U, B, V, R). Both spectra contain atmospheric absorption and the photomultiplier response function (see section \ref{filters} for details).}
    \label{band_filters}
    \end{center}
\end{figure}

As the spectrum of Cherenkov light (signal) and the night sky brightness (noise) are not identical, using filters to cut out long wavelengths may improve the signal to noise ratio. Figure \ref{band_filters} shows the photoelectron intensity at the PMT caused by NSB and by Cherenkov light in the four used photometric bands. The Cherenkov light has been simulated taking into account atmospheric absorption between the height of a typical air shower maximum and the observation altitude, and the PMT quantum efficiency. Both spectra are normalised for better comparison.

It can be seen that the Cherenkov light is strongest in U band and is constantly decreasing with wavelength, while the NSB peaks in the B band and does not fall off as quickly towards higher wavelengths as the Cherenkov light.

It is therefore reasonable to assume that the signal to noise ratio can be improved if long wavelengths are blocked, e.g.\ by a low-pass filter in front of the PMT. Table \ref {SN1} shows the fraction of Cherenkov light and NSB present in the various bands. If a filter blocks the V and R band but transmits the wavelengths in U and B band, about 92\% of the Cherenkov but only 57\% of the NSB will survive, resulting in an improvement of the signal to noise ratio of about 1.2. This requires however an ideal low-pass filter that reaches 100\% transmission at short wavelengths.

\begin{table}[t]
\begin{center}
\begin{tabular}{lccc}
\hline			
Included bands & signal & NSB & S / $\sqrt{\mathrm{NSB}}$ \\
\hline
U				& 0.55	& 0.21 &	1.20   \\
U, B			& 0.92  & 0.57 &	1.21   \\
U, B, V		& 1.00	& 0.92 &	1.04   \\
U, B, V, R	& 1.00	& 1.00 &	1.00   \\
\hline
\end{tabular}
\end{center}
\caption{Fraction of Cherenkov light and NSB present in the used photometric bands (using the same numbers as the plot in figure \ref{band_filters}). The last column shows the gain in the signal to noise ratio which is achieved if only the respective wavelength bands are transmitted to the PMT.
Shown are the numbers for dark regions as measured at Fowler's Gap; however, the numbers for bright regions are almost identical.}
\label{SN1}
\end{table}

\section{Implications for air Cherenkov detectors}
\label{implications}
\subsection{Cherenkov telescope system TenTen}
The current design for the TenTen telescope system assumes a mirror area of $\unit[23.8]{m^2}$ and a solid angle per pixel of $\unit[1.5 \times 10^{-5}]{sr}$ \citep{2008NIMPA.588...48R}. With these numbers, and assuming a light collection efficiency of 80\%, single pixel photon rates from $\unit[5.7 \times 10^{8}]{Hz}$ in dark regions and up to $\unit[2.0 \times 10^{9}]{Hz}$ in bright regions are expected. If the photomultipliers in the telescope camera are assumed to have a similar quantum efficiency as the one used in this study (see Eq. \ref{QE}), the expected photoelectron rates range from $\unit[53]{MHz}$ up to $\unit[190]{MHz}$. The numbers can increase additionally if a bright star is in the field of view of the particular pixel. These photoelectron rates are in the same regime as those for other Cherenkov telescope systems, e.g.\ H.E.S.S.\ typically sees a NSB photoelectron rate of $\unit[100]{MHz}$ \citep{2004APh....22..109A}. 

\subsection{Wide-angle Cherenkov detector HiSCORE}
The HiSCORE detector will consist of an array of non-imaging, angle integrating Cherenkov detectors. Every station will contain four detector modules, each consisting of a $\unit[20]{cm}$ diameter photomultiplier and a light collector (Winston cone) to increase the light sensitive area of the module. 
To derive the expected noise generated by NSB in one HiSCORE detector channel, the angular acceptance of the Winston cone is calculated using the simulation described in section \ref{solid_angle}. The upper opening is set at a diameter of $\unit[40]{cm}$, the half opening angle at $30^\circ$ and the reflectivity of the cone material at 90\%, according to current design plans for the detector \citep{2011arXiv1104.2336H}. The effective solid angle is calculated with Eq.\ \ref{solid_angle_eq} to be $\Omega = \unit[0.68]{sr}$. 

With these numbers, the photon rate is expected to be about $\unit[1.7 \times 10^{11}]{Hz}$ for the detector pointing towards dark regions. The respective photoelectron rate is $\unit[1.6 \times 10^{10}]{Hz}$ or 16 photoelectrons per nanosecond.

To calculate the highest expected NSB rates in one HiSCORE module, a light band with a photon intensity profile as found in the scan of the Galactic Centre region (figure \ref{GC_scan}) is used as light source for the ray-tracing simulation of the Winston Cone. This simulation gives a photon rate of $\unit[3.5 \times 10^{11}]{Hz}$, and a photoelectron rate of $\unit[3.3 \times 10^{10}]{Hz}$ or 33 photoelectrons per nanosecond. For comparison, a single bright star with magnitude zero adds less than 0.1 photoelectrons per nanosecond, so that even several bright stars within the field of view will not increase this level significantly.

The level of NSB determines the energy threshold of the detector system: First, the trigger threshold must be set high enough to keep the rate of false triggers from NSB fluctuations at a manageable level, second the accuracy of the reconstruction of events with low energy (and hence weak Cherenkov light emission) is limited by the signal to noise ratio. An upcoming paper by M.\ Tluczykont, D.\ Hampf, D.\ Horns et al.\ (2011) will present detailed simulations of the HiSCORE detector, including a discussion of the effects of NSB on the trigger threshold and the reconstruction.

\section{Conclusion and Outlook}
It can be concluded from the data that Fowler's Gap is an excellent astronomical location in respect to the night sky brightness. However, long term measurements of the NSB in combination with extinction measurements are needed to rate the quality of the site more reliably. Campaigns for more measurements at this and other sites in Australia are planned.

The night sky brightness varies greatly with the region of the sky looked at. Regions on the galactic plane are more than three times brighter than the darkest regions of the sky. However, plane scans show that these high light levels are restricted to within $b=\pm 10^\circ$ from the galactic plane, which means that the highest level of night sky brightness seen by the wide-angle Cherenkov detector HiSCORE will be at most twice that of dark regions.

The night sky brightness is much stronger at longer wavelengths. However, standard photomultiplier sensitivities peak at short wavelengths, so that the most dominant contribution for the background is from the B and V bands. The peak in Cherenkov light is at shorter wavelengths, so that filters can be employed to improve the signal to noise ratio up to a factor of about 1.2.

\section*{Acknowledgements}
The authors thank the School of Chemistry \& Physics, University of Adelaide, for funding support.

Daniel Hampf likes to thank the German Federal Ministry of Education and Research (BMBF) for its financial support (contract number  05A08GU1), as well as the German Academic Exchange Service (DAAD) for its support of the field studies in Australia under the ''Kurzstipendium für Doktoranden`` scheme.

\bibliographystyle{elsarticle-harv}
\bibliography{NSB_paper}

\end{document}